\documentstyle[12pt,psfig]{article}
\begin{document}
\begin{center}
\Large{\bf Production and Thermal Equilibration of Partons\\
 in\\
 Relativistic Heavy Ion
Collisions:\\
 $\sqrt{s}=$ 20 {\em vs.} 200 A$\cdot$GeV} 
\vskip 0.2in

\large{Dinesh Kumar Srivastava }
\vskip 0.2in

\large{\em Variable Energy Cyclotron Centre,\\
 1/AF Bidhan Nagar, Calcutta 
700 064, India\\
and\\
Fakult\"{a}t f\"{u}r Physik, Universit\"{a}t Bielefeld, D-33501,
Bielefeld, Germany}

\vskip 0.2in

Abstract

\vskip 0.2in

\end{center}
The production and thermalization of partons liberated in relativistic
heavy ion collisions is investigated
 within the parton cascade model. The momentum distribution of the partons
near $z=0$
is found to become isotropic beyond about 1 fm/$c$ after the overlap for
the collision of lead nuclei at $\sqrt{s}=$20 A$\cdot$GeV
indicating a thermalization. At 
$\sqrt{s}=$200 A$\cdot$GeV this is attained by about 0.5 fm/$c$ after
the complete overlap.  

\section{Introduction}

The search for quark-gluon plasma, the deconfined  strongly 
interacting matter, has entered a decisive phase with the
 relativistic heavy ion collision experiments  slated to begin within a
few months
at the Brookhaven National Laboratory. This will put the
scores of theoretical models, conjectures, and speculations put forward over
the last decade as plausible signatures of quark gluon plasma 
(QGP)~\cite{jaipur} to a severe test.
The corresponding search at the CERN SPS has already yielded a large body
of data which have been carefully examined for the evidence of QGP. 
It has been suggested that it is quite likely that the partonic phase
may, indeed, have been reached in sulfur and lead induced collisions
at the CERN SPS~\cite{stock}.

The parton cascade model proposed by Geiger and M\"{u}ller~\cite{gm} and
refined and studied exhaustively by Geiger and coworkers~\cite{all}
attempts to describe the relativistic collision of nuclei using the
partonic picture of hadronic interactions where the nuclear dynamics is
described in terms of quark and gluon interactions within perturbative
QCD, embedded in the frame work of relativistic transport theory. 
The complete space-time picture of the evolution is simulated by solving
an appropriate transport equation in six-dimensional phase space using
Monte Carlo techniques. The model supplemented with a cluster
hadronization scheme has been developed into a computer code 
VNI~\cite{vni}. It was recently shown to give a reasonable description 
to particle spectra at SPS energies for $S+S$ and $Pb+Pb$ collisions~\cite{sps},
which needs to be understood in a greater detail in view of
the suggestion made in Ref.~\cite{stock}.

In the present work we continue this effort~\cite{other} and use the power of
 the parton cascade
model to see how quickly does the partonic matter produced in such
collisions thermalize at CERN SPS and BNL RHIC energies. It should be added
that this question has been studied within the context of RHIC (and
LHC) energies by Geiger~\cite{all}, and we display them somewhat
differently (and with much better statistics) for a ready and
easy comparison.

Even though the developments in the parton cascade model have been
very well documented, it is worthwhile recalling the most important
steps:
\begin{itemize}
\item
The {\em initial state} associated with the incoming nuclei involves
their decomposition into nucleons and of the nucleons into partons
on the basis of experimentally measured nucleon structure
functions and elastic form factors. This procedure then translates the
initial nucleus-nucleus system into two colliding clouds of virtual partons.

\item
 The {\em parton cascade development} starts from the initial
inter-penetrating parton clouds and traces their space-time development
with mutual interactions and self interactions of the system of quarks and
gluons. The model includes {\em multiple} elastic and inelastic 
interactions described as sequences of elementary $2\,\rightarrow \,2$
scatterings, $1\, \rightarrow \, 2$ emissions, and $2 \, \rightarrow \,1$
fusions. Several important effects which characterize the space-time
evolution of a many parton system in nuclear collisions like the
individual time scale of each parton-parton collision, the formation
time of the parton radiation, the effective suppression of radiative
emissions from virtual partons due to an enhanced absorption probability
of others in regions of dense phase space occupation, and the
effects of soft gluon interference in low energy gluon emissions
are explicitly accounted for.

\item 
And finally, the {\em hadronization dynamics} of the evolving system in
terms of a parton coalescence to colour neutral clusters is described as
a local statistical process that depends on the spatial separation
and colour of the nearest-neighbour partons~\cite{eg}. 
The pre-hadronic clusters then decay to form hadrons.

\end{itemize}

In the present paper we shall be concerned with only up to the
second stage of the collision, for times $\leq $ 2 fm/$c$ so that  the matter
is still in the form of primary, secondary, and space-like partons
and which essentially spans the so-called pre-equilibrium era of
the evolution.

We consider four cases: $S+S$  and $Pb+Pb$ collisions at $\sqrt{s}=$ 
20 A$\cdot$GeV and 200 A$\cdot$GeV. In all the cases we place the
two nuclei centred at $z=\pm 1$ fm at $t=-$ 1 fm/$c$, which will propel
them on to a complete overlap at around $t=0$ fm/$c$.

 We analyze the results for the time evolution of the longitudinal and
rapidity distribution of partons and also see how the thermalization
is attained and maintained, if at all.

It should be added right at the outset that only the `real' partons have
been included in these spectra~\cite{gm}. Thus the initial state, before
the collision will reflect only the distribution of valence quarks,
and as the collision proceeds more and more of the (initially)
space-like gluons and sea-quarks gain enough energy to become either
time-like or be on the mass-shell. One may also add that the rapidity
variable is not defined for space-like particles (as for them $E<|p_z|$)
and moreover, the partons which remain space-like throughout the
collision do not contribute to the reaction dynamics and will be
reabsorbed during the hadronization.

\section{$\sqrt{s}=$ 20 A$\cdot$GeV}

\subsection{Production of (semi)hard partons}

In Fig.~1, we have shown the time development of the longitudinal
distribution of (real) partons for $S+S$ collision at 20 A$\cdot$GeV.
 We see that the the partonic distributions almost touch when
$t=-0.4$ fm/$c$ and the nuclei  disengage by $t\approx$ 1 fm/$c$, 
leaving a trail of  secondary partons near $z=0$, which are created 
in the (semi)hard collisions and radiations.

The evolution of the rapidity distribution of the partons  for $S+S$
collisions is shown in Fig.~2. 
We see that initially they are distributed over
about four units of rapidity, and peak around $y=\pm$ 2.
 As a result of the collision the number of partons having 
$y \approx \pm $ 2 gets reduced, and most of the secondary partons 
materialize by the time $t=$ 0.4 fm/$c$. They are seen to be confined to 
$|y|\leq$ 2.  Approximating the width of
 the region over which the secondary partons are spread,
as $\Delta y \approx $ 4, we see that there is a production of about
 two partons per participating nucleon.
 The dip in the rapidity distribution is quite interesting too, as it implies
that a large part of the partons have just passed through, without
interacting.

The situation for the collision of lead nuclei is quite
different. We see (Fig.~3) that the nuclei start touching at $t=-$ 0.8
fm/$c$ and there is already a considerable overlap by the time $t=-$ 0.4 fm/$c$.
The nuclei disengage only by $t\approx$ 1.5 fm/$c$, leaving a trail of
secondary partons in their wake. 

The differences between the two cases are much more dramatic, when
we look at the evolution of the rapidity distribution of the partons
(Fig.~4).
Thus we see that there is a considerable production of partons already
at $t=0$ fm/$c$, even when the nuclei have not yet ploughed through
each other. The particle production continues till about 1.2 fm/$c$
which is clear from the  modification of the rapidity distribution till then.
We also see that now the parton production is much more, the dip at
$y=0$ in the rapidity distribution before the collision is completely
filled up by the end of the collision, and we have a flat top
distribution of partons!
We also see that again the secondary partons are spread over $|y|\leq$ 2  
and up to four partons per participating nucleon are produced.
Recalling that the nuclei would be thicker at smaller transverse
distances from the collision axis and that nucleons there are
more likely to re-scatter, these observations imply a considerable 
multiple scattering among the partons.

\subsection{Thermalization of partons}

As indicated in the Introduction, the thermalization of partons has been
addressed in detail in the parton cascade model studies. Several
phenomenological estimates have also been obtained in the 
literature~\cite{thermal}. We adopt a 
slightly different strategy here and look at the $|p_x|, \ |p_y|,$
and $|p_z|$ distribution of the materialized partons at different times
in the centre of mass system of the colliding nuclei. We confine our
attention to $|z|\, \leq$ 0.5 fm, and determine whether these
distribution become isotropic at some stage during the evolution. A
similar approach was used in Ref.~\cite{kari} when only the radiation
of the gluons following the first (semi)hard scatterings was included in
such collisions. However, it is expected that multiple scatterings
included in the parton cascade model used in the
present work will hasten this process and also maintain
the thermal equilibrium. In absence of the multiple scatterings a
thermal equilibrium can neither be achieved nor maintained.

The results of our simulations for $S+S$ collisions at $\sqrt{s}$ =20
A$\cdot$GeV are shown in Fig.~5. We see that the scatterings which 
began even before $t=0$ fm, (see Figs. 1 \& 2) increase the value of
$<|p_x|>$ and $<|p_y|>$ and decreases the $<|p_z|>$, as seen from the
reduction of the number of partons having large $|p_z|$ and their 
reappearance with
smaller $|p_z|$. There will also be an escape of partons  which
have a large rapidity from the zone $|z|\leq$ 0.5 fm chosen here. As
a result of these two processes the momentum distribution becomes
isotropic around $t=$ 1 fm/$c$, and stays so for a brief while after 
that.

The corresponding results for $Pb+Pb$ collisions (Fig.~6) are similar in
nature and we see that the number of partons released is much
larger in the heavier system. Isotropy of the momenta is attained
in the system near $z=0$ around $t=$ 1 fm/$c$ which is maintained
afterwards.

 Thus we conclude that the partonic system produced in heavy-ion
collisions at $\sqrt{s}=$ 20 A$\cdot$GeV
will have an excursion into a thermalized zone at about 1 fm/$c$
after the nuclei overlap as a result of (semi)hard partonic collisions
and gluonic radiations. 

\section{$\sqrt{s}=$ 200 A$\cdot$GeV}

While the attempt to use the parton cascade model at such low energies
as  $\sqrt{s}=$ 20 A$\cdot$GeV are rather recent, those at the energies
relevant to BNL RHIC, 200 A$\cdot$GeV, are quite well documented and
thus we shall give only the results for the rapidity distribution and
the approach to isotropic momenta in the central zone of the collision
volume.

\subsection{Production of (semi)hard partons}

Thus in Fig.~7 we have shown the time evolution of the rapidity
distribution of the real partons for central collisions of sulfur
nuclei. As before the dashed histograms 
show the results for the primary (uninteracted) partons while the
solid histograms show the sum of the primary and the secondary
(semi)hard partons produced in the collision. We note that the 
production of partons in the parton cascade approach utilized here
is completed by 0.4 fm/$c$ after the nuclei overlap fully.
We can also estimate that the up to 6 partons per `participating'
nucleon may be produced in these collisions which are spread over 
$|y|\leq$ 2.5.

The corresponding results for central collisions of lead nuclei
(Fig.~8) are very revealing, as the dip in the rapidity distribution
of the partons is much less pronounced, as a result of a much larger
production of partons due to multiple scatterings. We also estimate 
that up to 9 partons per nucleon may be produced in this case.

Before moving on, it is of interest to note that the shape of
the initial state parton distribution here is different from the
one in Fig.~2 (at 20 A$\cdot$GeV) as we prepare the assembly of
the partons at a higher $Q_0^2$ (see, Discussion), resulting in
a higher number of partons at the lower end of $x$, which translates into
smaller $|y|$ in this plot.

\subsection{Thermalization of partons}

The evolution of the momentum distribution of the partons produced due
to the (semi)hard scatterings and radiations are shown in Fig.~9 and ~10
respectively for the central collisions of sulfur and lead nuclei 
at energies relevant to BNL RHIC.
We see that initially the partons have a large $|p_z|$, (note also the dip
at $|p_z|\approx $ 0, depicting the separation of the partons
in the rapidity space) and the collisions transfer $p_z$ into $p_x$
and $p_y$ which increase substantially and rapidly. The peak in the $|p_z|$
spectrum around 1 GeV holds the primary partons which have not
interacted yet. We note that by the time $t=$ 0.4 fm/$c$ the slopes
of the momenta along the three direction are similar, but for the above
mentioned peak, which persists till the end. Ignoring these uninteracted
partons, we see that the partons which materialize during the collision
attain an isotropic momentum distribution by $t\approx$ 0.5 fm/$c$.
 Once again, the
partons having large rapidities will escape the longitudinal slice which
we have considered.

\section{Discussion}

Before concluding it is worthwhile that some aspects of the model used
for the studies reported here are reiterated. The first one concerns
the so-called $p_0$ which is used to divide the scatterings in the
parton cascade model into soft (elastic) and (semi)hard  reactions.
 It is taken as
1.12 GeV for the collisions at 20 A$\cdot$GeV  and 2.09 GeV for
 collisions at 200 A$\cdot$GeV based on
considerations of $pp$ cross-sections discussed in detail~\cite{gm,all}. 
There is no reason to believe that they should have the same values for
nuclear collisions, but these provide a convenient starting point. 
An increase in $p_0$ will obviously decrease the (semi)hard scatterings.
The scale $Q_0$ at which the nucleon structure functions are
initialized are chosen as 1.32 and 2.35 GeV at the two energies, based
on an estimate of $<p_T^2>$ in primary-primary collisions among the
primary partons. We have already remarked that a larger $Q_0$ 
leads to the different shapes for the rapidity distributions at the higher
energy considered here.
 
A careful reader must have noticed that the transverse components of
the momenta ``stabilize'' fairly quickly. This has
its origin in the fact that {\em no} collisions between partons
is permitted,
if their relative $\sqrt{s}$ is less than 2 GeV. This is done to
ensure that the perturbative QCD treatment used in the parton cascade
model remains valid. Again, a parton created in a scattering with a 
large momentum looses energy quickly due to gluonic radiations
 (till its virtuality drops to the cut-off $\mu_0\,\approx$ 1 GeV chosen
in the calculations) and its
momentum is considerably reduced before it undergoes next collision.
When the initial energy is high and the parton density is large, it may
participate in further hard scatterings with production of additional
partons, else- it will undergo only soft scatterings (see, e.g.,
 Ref.~\cite{other,mik}). 

 A more comprehensive~\cite{bm} approach could proceed as follows: use the 
(practical) procedure of soft and (semi)hard scatterings of the parton 
cascade model to initiate the collision
and once the partonic density is large enough to ensure screening of the
long range forces on a scale where the Debye mass $\mu_D$ is much larger
than $\Lambda_{QCD}$, remove the  division of soft and hard scatterings
implemented in the model; evaluating the scattering in terms of the
Debye mass  at {\em all}  possible $\sqrt{s}$ between partons and for
 {\em all}
momentum transfers. This treatment will, however, quickly get beyond the
capability of most of the computers due to the very large number
of scatterings taking place. The other and more serious problem will involve
the comparatively small number of partons in a given event to implement
this scheme over the entire space-time spanned by the system as  then we
would be plagued by large fluctuations.
 A more practical approach~\cite{sspc} could
be to invoke hydrodynamics to pursue the evolution~\cite{smm}
 beyond the point of thermalization, after an average is taken over a
large number of events. This is in progress.

\section{Conclusions}

To conclude, we have seen that the parton cascade model suggests that
there is a substantial production of partonic matter as a result of
 (semi)hard scatterings and QCD branchings in 
$S+S$ and $Pb+Pb$ collisions at 20 A$\cdot$GeV, which may attain
an isotropy in momentum distribution at about 1 fm/$c$ after the
nuclei overlap completely at $|z|\approx 0$. The simulations reveal 
an increased
multiple scattering activity in lead induced collisions.

The corresponding production at 200 A$\cdot$GeV is (obviously) much
larger and the isotropy in the momentum distribution is attained
within 0.5 fm/$c$. It is also suggested that this could be
a convenient point to initialize a hydrodynamic approach to the
evolution, if desired.

\section*{Acknowledgments} 
The author gratefully acknowledges the hospitality of University of
Bielefeld where part of this work was done. He would also like to acknowledge
useful comments from Hans Gutbrod and Bikash Sinha.

\newpage

\centerline {FIGURE CAPTIONS}

\vspace{0.2in}

\noindent{Fig.~1  Longitudinal distribution of (real) partons
in central collision of sulfur nuclei at $\sqrt{s}=$ 20 A$\cdot$GeV,
at different times before and after the collision.
The solid histograms give the sum of primary and (semi)hard secondary partons,
while the dashed histograms give the result for the
primary (uninteracted) partons.
}

\vspace{0.2in}

\noindent{Fig.~2  Rapidity distribution of (real) partons
in central collision of sulfur nuclei at $\sqrt{s}=$ 20 A$\cdot$GeV,
at different times before and after the collision.
The solid histograms give the sum of primary and (semi)hard secondary partons,
while the dashed histograms give the result for the
primary (uninteracted) partons.
}
\vspace{0.2in}

\noindent{Fig.~3  Longitudinal distribution of (real) partons
in central collision of lead nuclei at $\sqrt{s}=$ 20 A$\cdot$GeV,
at different times before and after the collision.
The solid histograms give the sum of primary and (semi)hard secondary partons,
while the dashed histograms give the result for the
primary (uninteracted) partons.
}
\vspace{0.2in}

\noindent{
Fig.~4  Rapidity distribution of (real) partons
in central collision of lead nuclei at $\sqrt{s}=$ 20 A$\cdot$GeV,
at different times before and after the collision.
The solid histograms give the sum of primary and (semi)hard secondary partons,
while the dashed histograms give the result for the
primary (uninteracted) partons.
}

\vspace{0.2in}

\noindent{
Fig.~5 Time evolution of the $|p_x|$ (crosses), $|p_y|$ (diamonds)
and $|p_z|$ (histogram) distribution of the (real) partons in 
central collision of sulfur nuclei at $\sqrt{s}=$  20 A$\cdot$GeV.
}
\vspace{0.2in}

\noindent{
Fig.~6 Time evolution of the $|p_x|$ (crosses), $|p_y|$ (diamonds)
and $|p_z|$ (histogram) distribution of the (real) partons in 
central collision of lead nuclei at $\sqrt{s}=$  20 A$\cdot$GeV.
}
\vspace{0.2in}

\noindent{Fig.~7  Rapidity distribution of (real) partons
in central collision of sulfur nuclei at $\sqrt{s}=$ 200 A$\cdot$GeV,
at different times before and after the collision.
The solid histograms give the sum of primary and (semi)hard secondary partons,
while the dashed histograms give the result for the
primary (uninteracted) partons.
}
\vspace{0.2in}

\noindent{Fig.~8  Rapidity distribution of (real) partons
in central collision of lead nuclei at $\sqrt{s}=$ 200 A$\cdot$GeV,
at different times before and after the collision.
The solid histograms give the sum of primary and (semi)hard secondary partons,
while the dashed histograms give the result for the
primary (uninteracted) partons.
}
\vspace{0.2in}

\noindent{
Fig.~9 Time evolution of the $|p_x|$ (crosses), $|p_y|$ (diamonds)
and $|p_z|$ (histogram) distribution of the (real) partons in 
central collision of sulfur nuclei at $\sqrt{s}=$  200 A$\cdot$GeV.
}
\vspace{0.2in}

\noindent{
Fig.~10 Time evolution of the $|p_x|$ (crosses), $|p_y|$ (diamonds)
and $|p_z|$ (histogram) distribution of the (real) partons in 
central collision of lead nuclei at $\sqrt{s}=$  200 A$\cdot$GeV.
}
\vspace{0.2in}
\newpage

\begin{figure}
\psfig{file=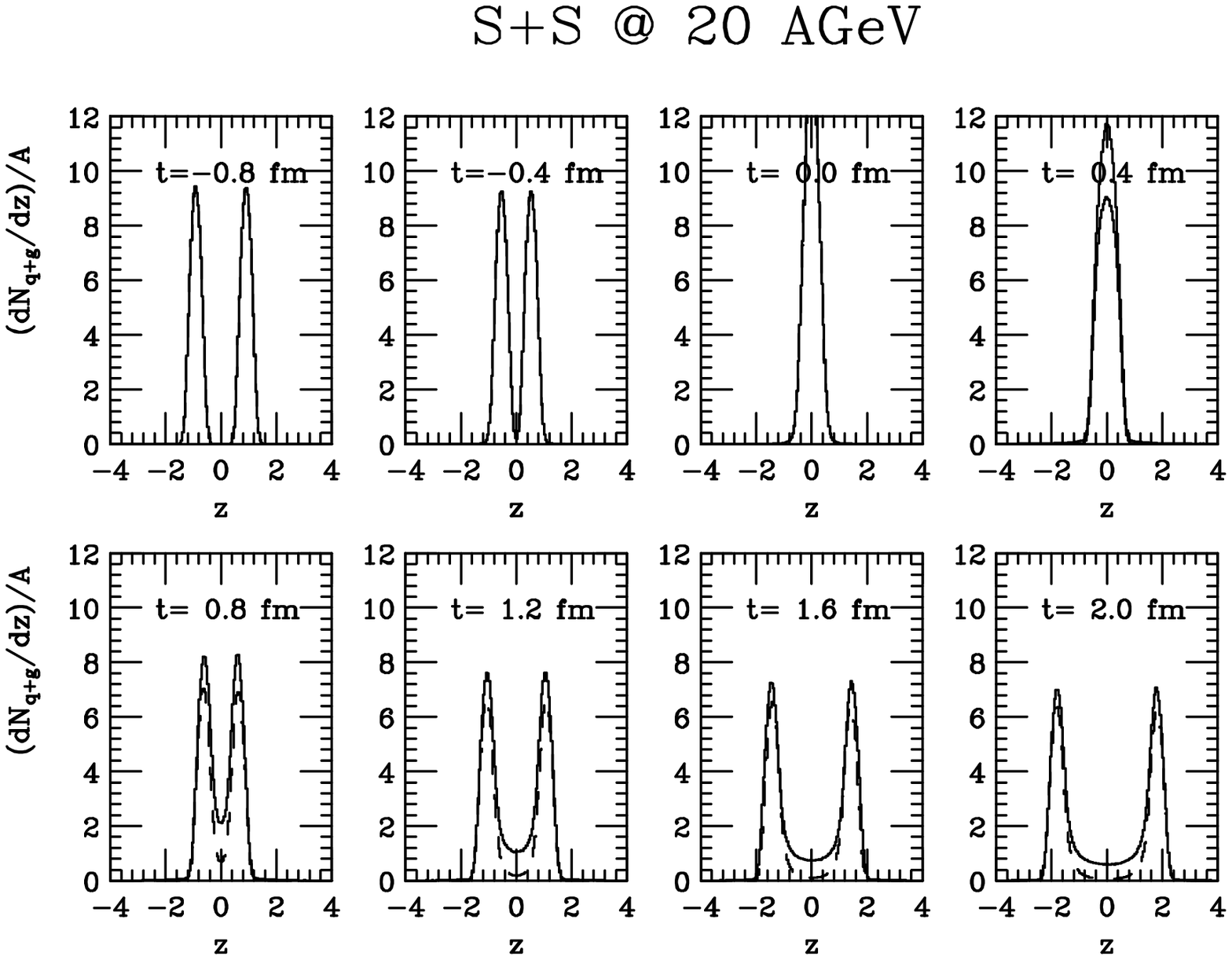,height=12cm,width=15cm}
\vskip 0.1in
\caption{
 Longitudinal distribution of (real) partons
in central collision of sulfur nuclei at $\sqrt{s}=$ 20 A$\cdot$GeV,
at different times before and after the collision.
The solid histograms give the sum of primary and (semi)hard secondary partons,
while the dashed histograms give the result for the
primary (uninteracted) partons.
}
\end{figure}

\newpage

\begin{figure}
\psfig{file=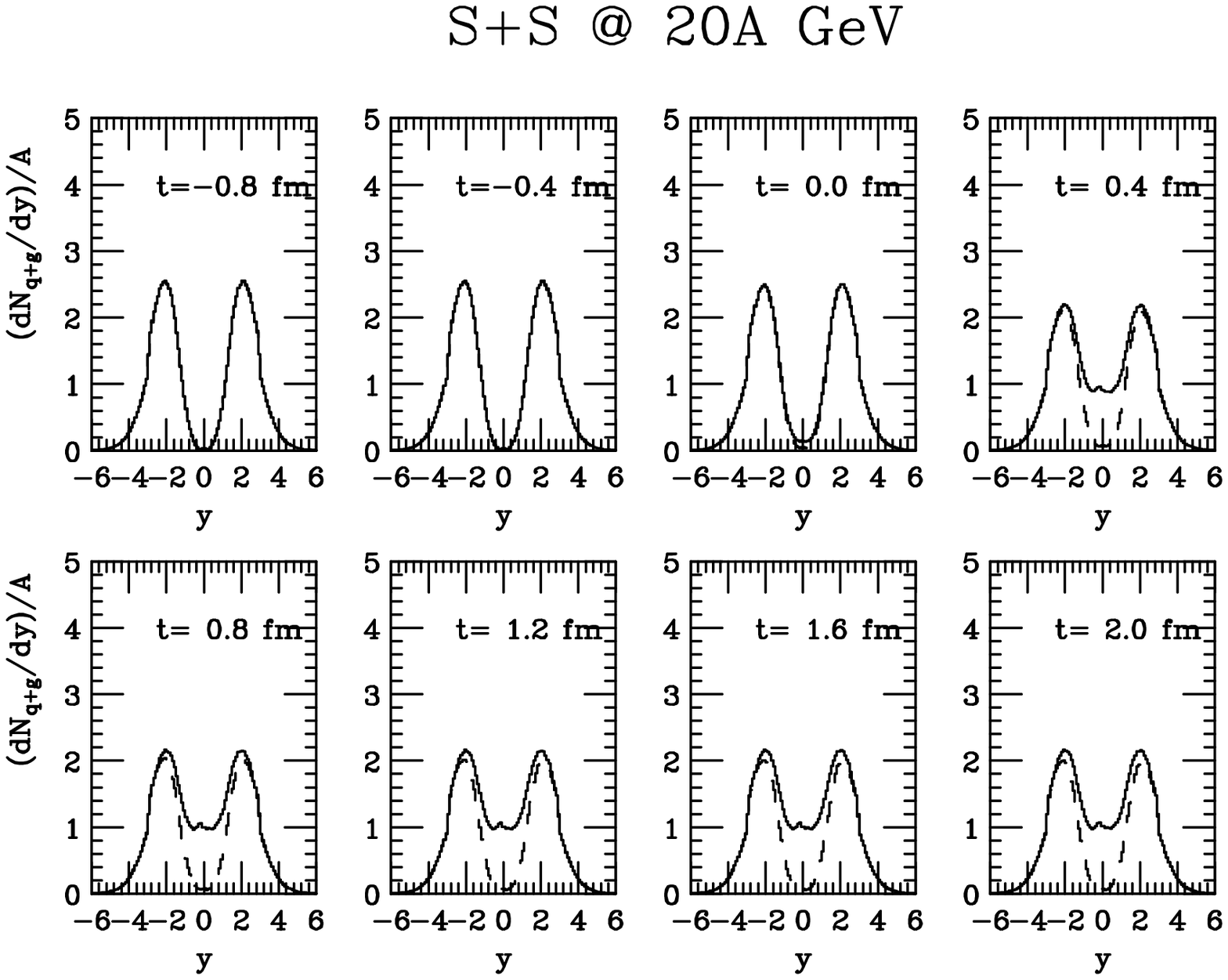,height=12cm,width=15cm}
\vskip 0.1in
\caption{
  Rapidity distribution of (real) partons
in central collision of sulfur nuclei at $\sqrt{s}=$ 20 A$\cdot$GeV,
at different times before and after the collision.
The solid histograms give the sum of primary and (semi)hard secondary partons,
while the dashed histograms give the result for the
primary (uninteracted) partons.
}
\end{figure}
\newpage

\begin{figure}
\psfig{file=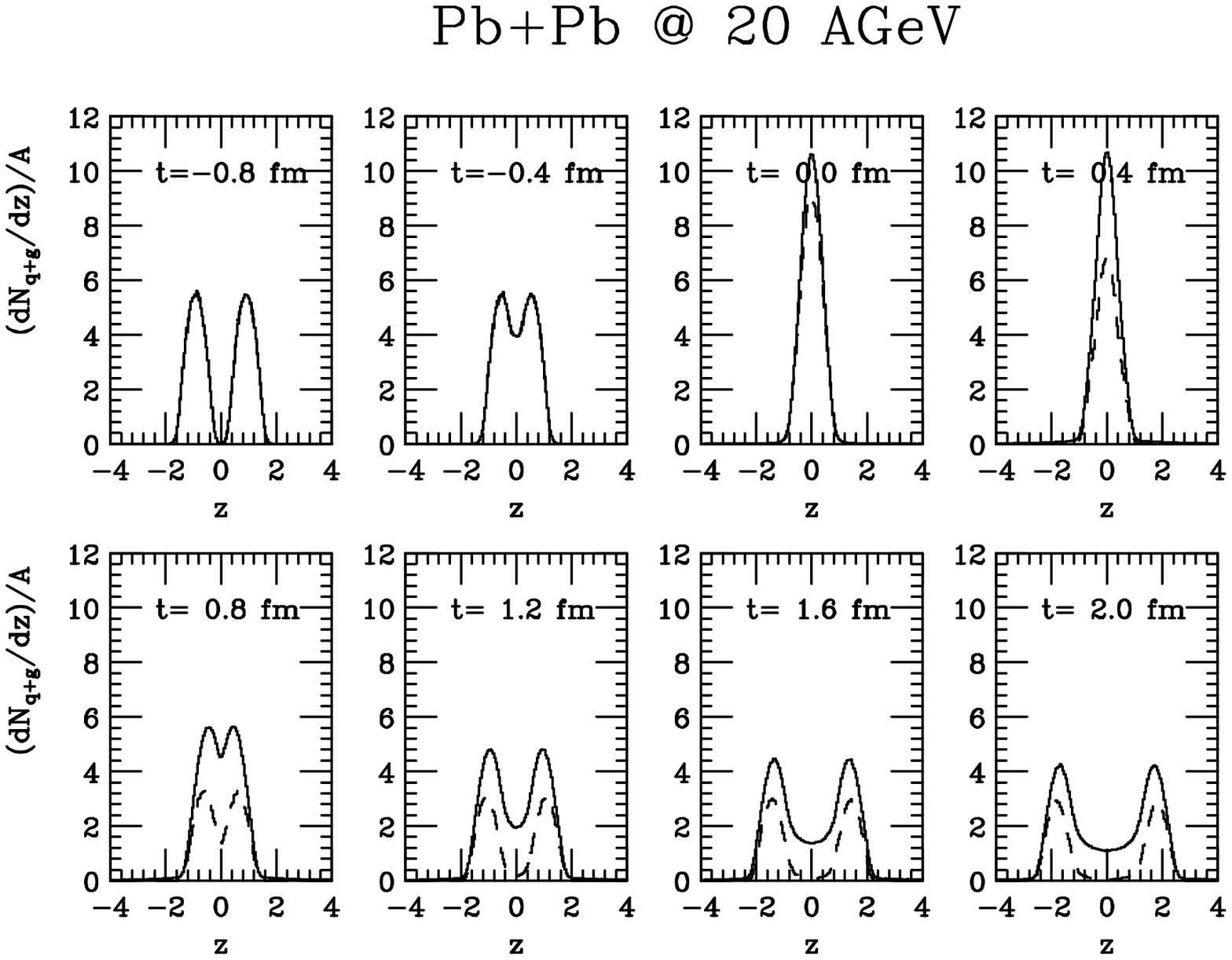,height=12cm,width=15cm}
\vskip 0.1in
\caption{
 Longitudinal distribution of (real) partons
in central collision of lead nuclei at $\sqrt{s}=$ 20 A$\cdot$GeV,
at different times before and after the collision.
The solid histograms give the sum of primary and (semi)hard secondary partons,
while the dashed histograms give the result for the
primary (uninteracted) partons.
}
\end{figure}

\newpage

\begin{figure}
\psfig{file=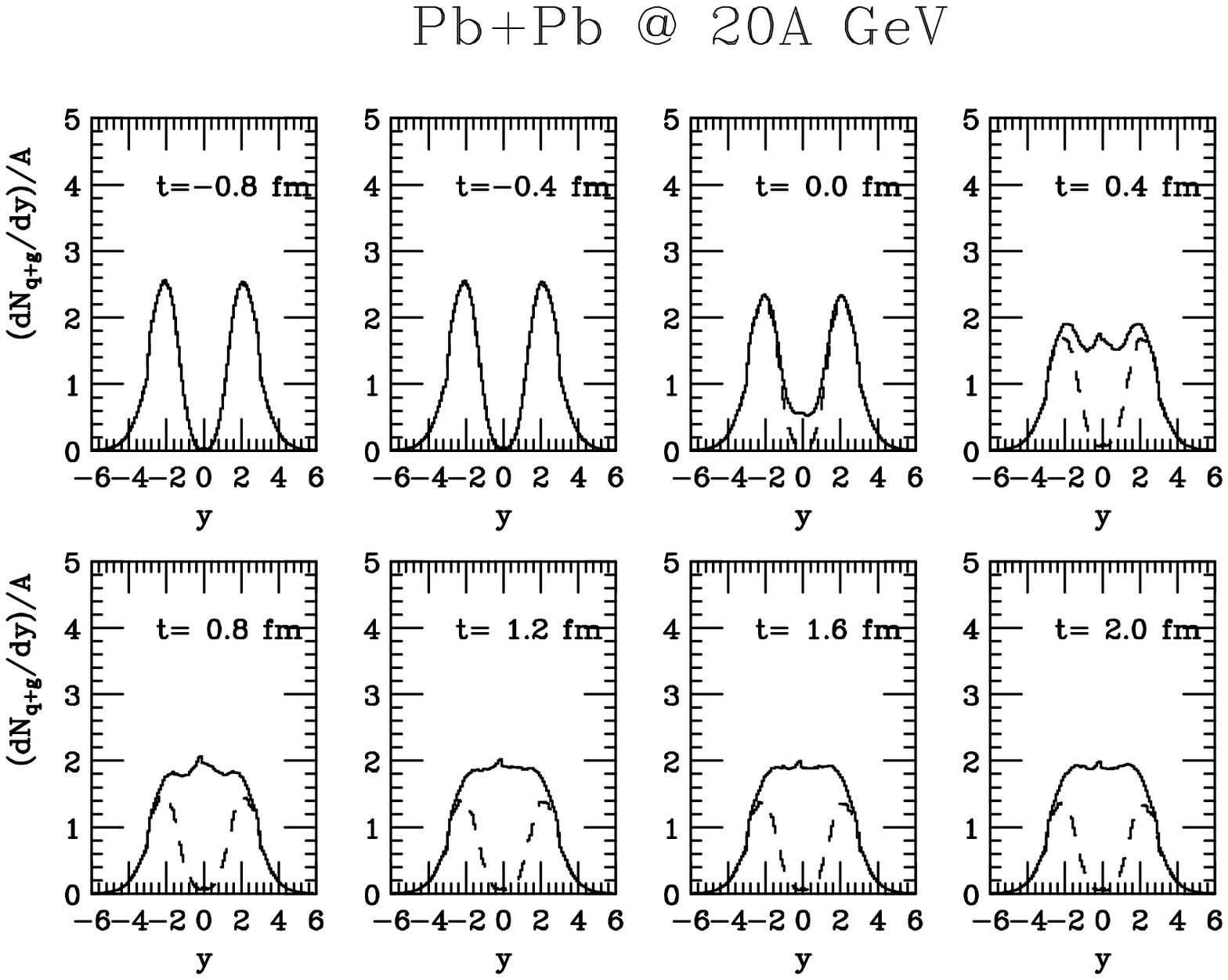,height=12cm,width=15cm}
\vskip 0.1in
\caption{
  Rapidity distribution of (real) partons
in central collision of lead nuclei at $\sqrt{s}=$ 20 A$\cdot$GeV,
at different times before and after the collision.
The solid histograms give the sum of primary and (semi)hard secondary partons,
while the dashed histograms give the result for the
primary (uninteracted) partons.
}
\end{figure}

\newpage

\begin{figure}
\psfig{file=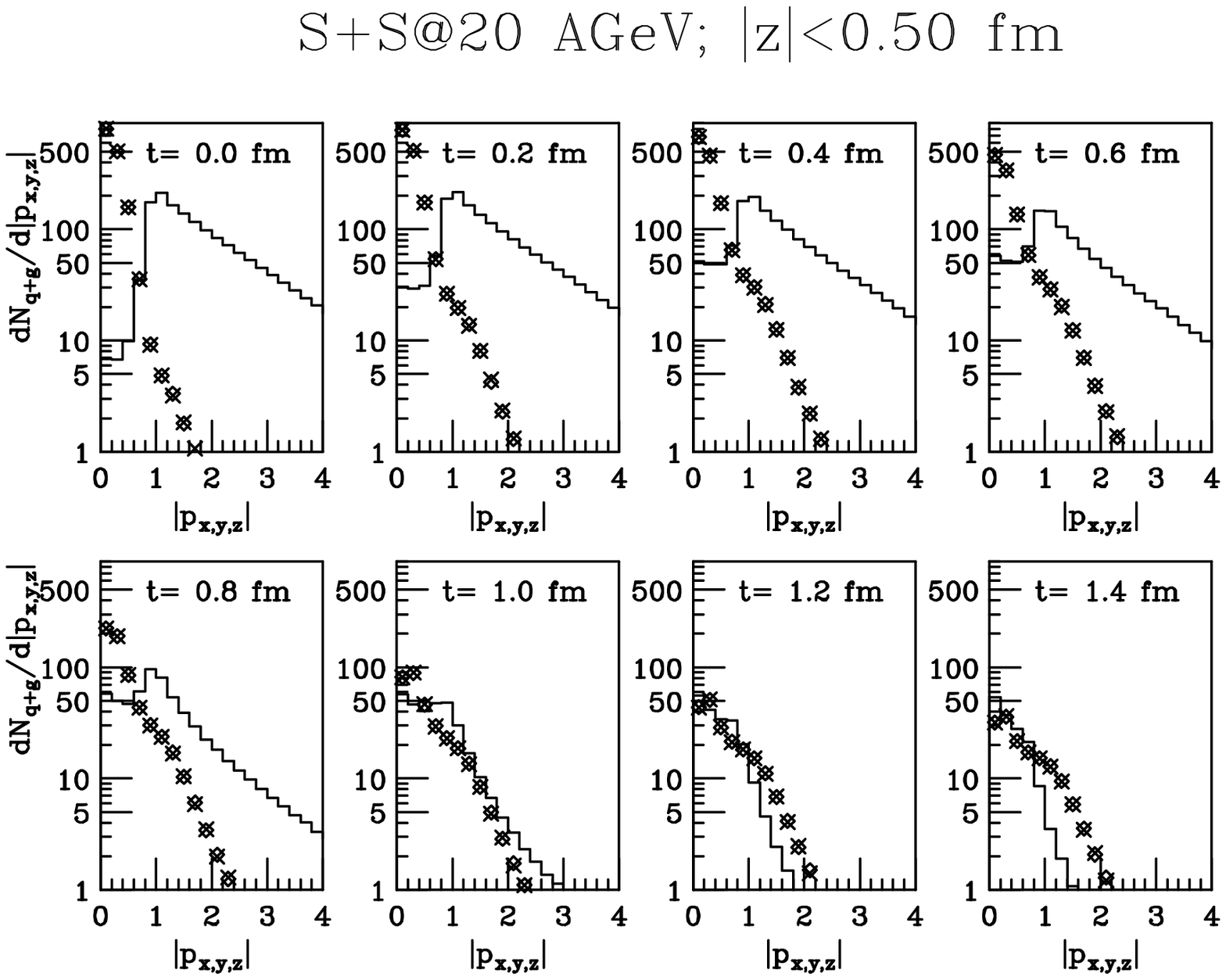,height=12cm,width=15cm}
\vskip 0.1in
\caption{
        Time evolution of the $|p_x|$ (crosses), $|p_y|$ (diamonds)
and $|p_z|$ (histogram) distribution of the (real) partons in 
central collision of sulfur nuclei at $\sqrt{s}=$  20 A$\cdot$GeV.
}
\end{figure}

\newpage

\begin{figure}
\psfig{file=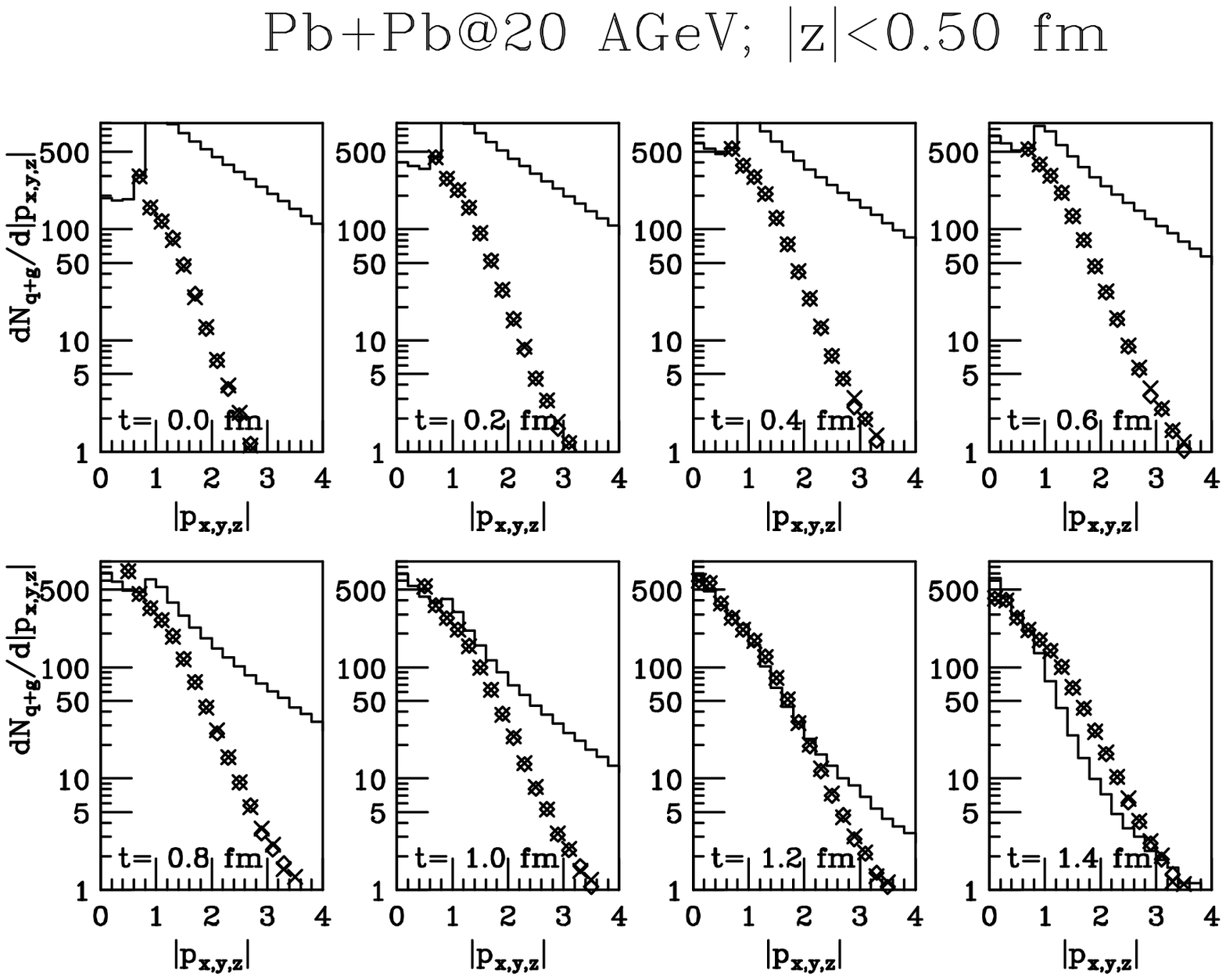,height=12cm,width=15cm}
\vskip 0.1in
\caption{
        Time evolution of the $|p_x|$ (crosses), $|p_y|$ (diamonds)
and $|p_z|$ (histogram) distribution of the (real) partons in 
central collision of lead nuclei at $\sqrt{s}=$  20 A$\cdot$GeV.
}
\end{figure}
\newpage
\begin{figure}
\psfig{file=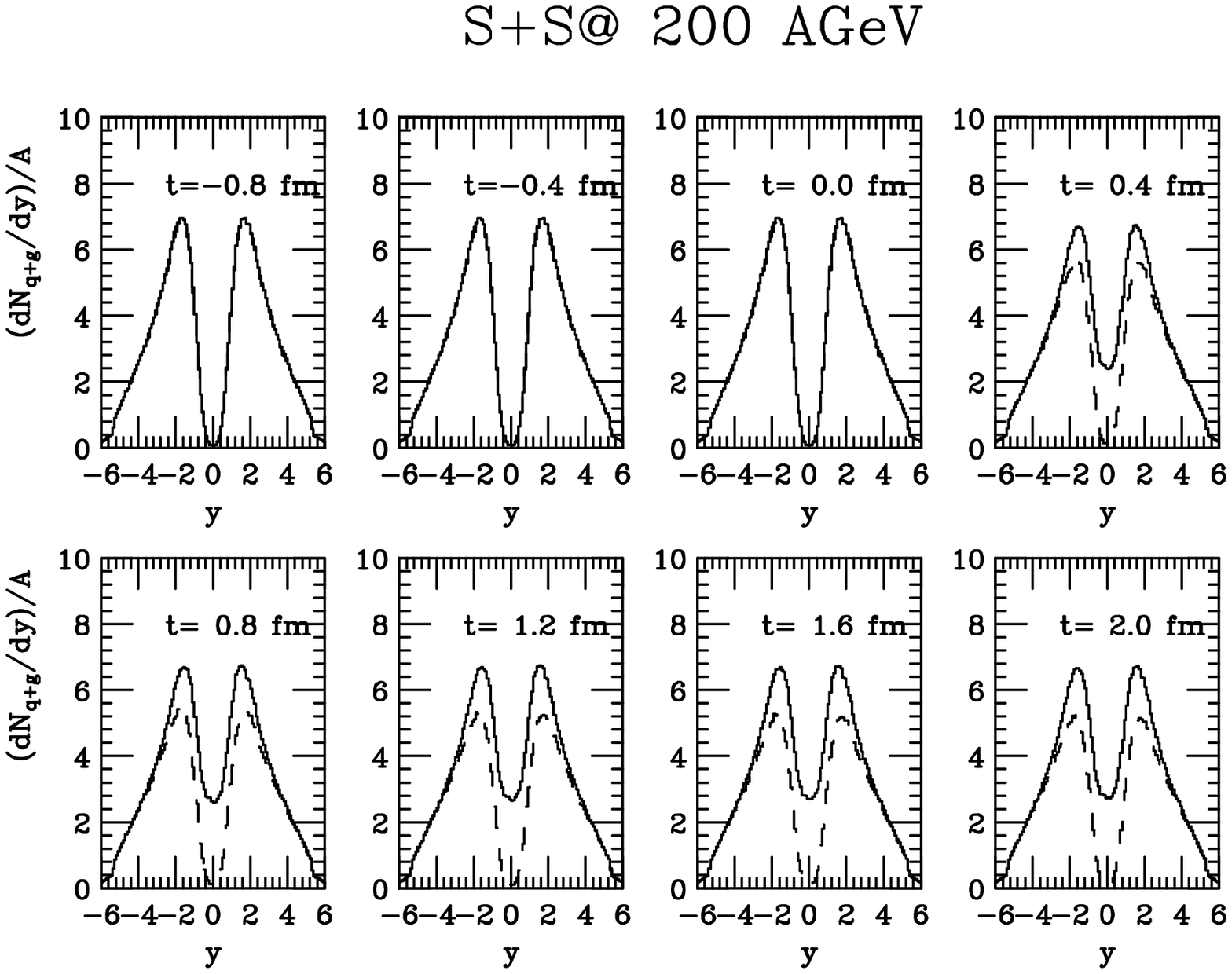,height=12cm,width=15cm}
\vskip 0.1in
\caption{
  Rapidity distribution of (real) partons
in central collision of sulfur nuclei at $\sqrt{s}=$ 200 A$\cdot$GeV,
at different times before and after the collision.
The solid histograms give the sum of primary and (semi)hard secondary partons,
while the dashed histograms give the result for the
primary (uninteracted) partons.
}
\end{figure}
\newpage
\begin{figure}
\psfig{file=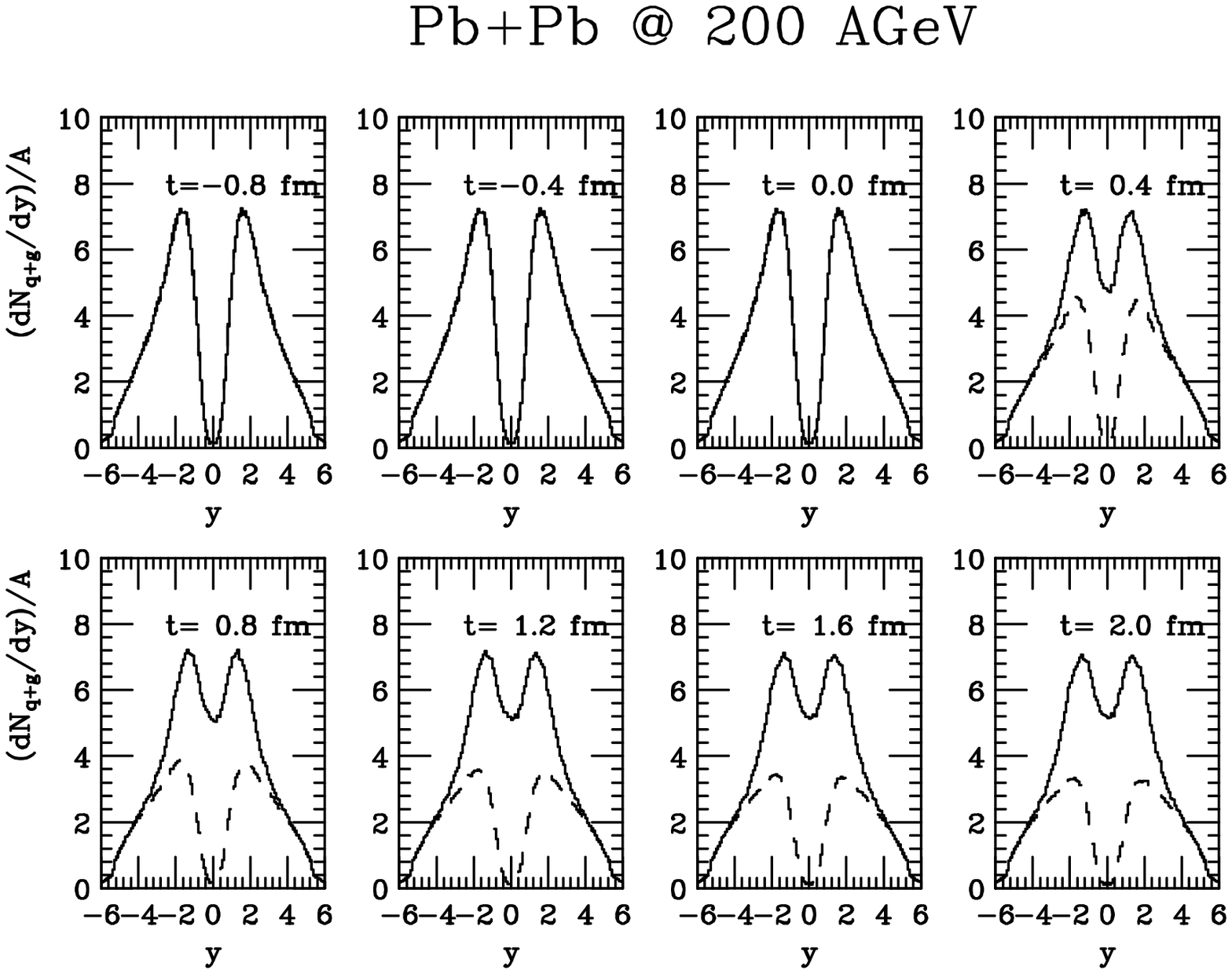,height=12cm,width=15cm}
\vskip 0.1in
\caption{
  Rapidity distribution of (real) partons
in central collision of lead nuclei at $\sqrt{s}=$ 200 A$\cdot$GeV,
at different times before and after the collision.
The solid histograms give the sum of primary and (semi)hard secondary partons,
while the dashed histograms give the result for the
primary (uninteracted) partons.
}
\end{figure}
\newpage
\begin{figure}
\psfig{file=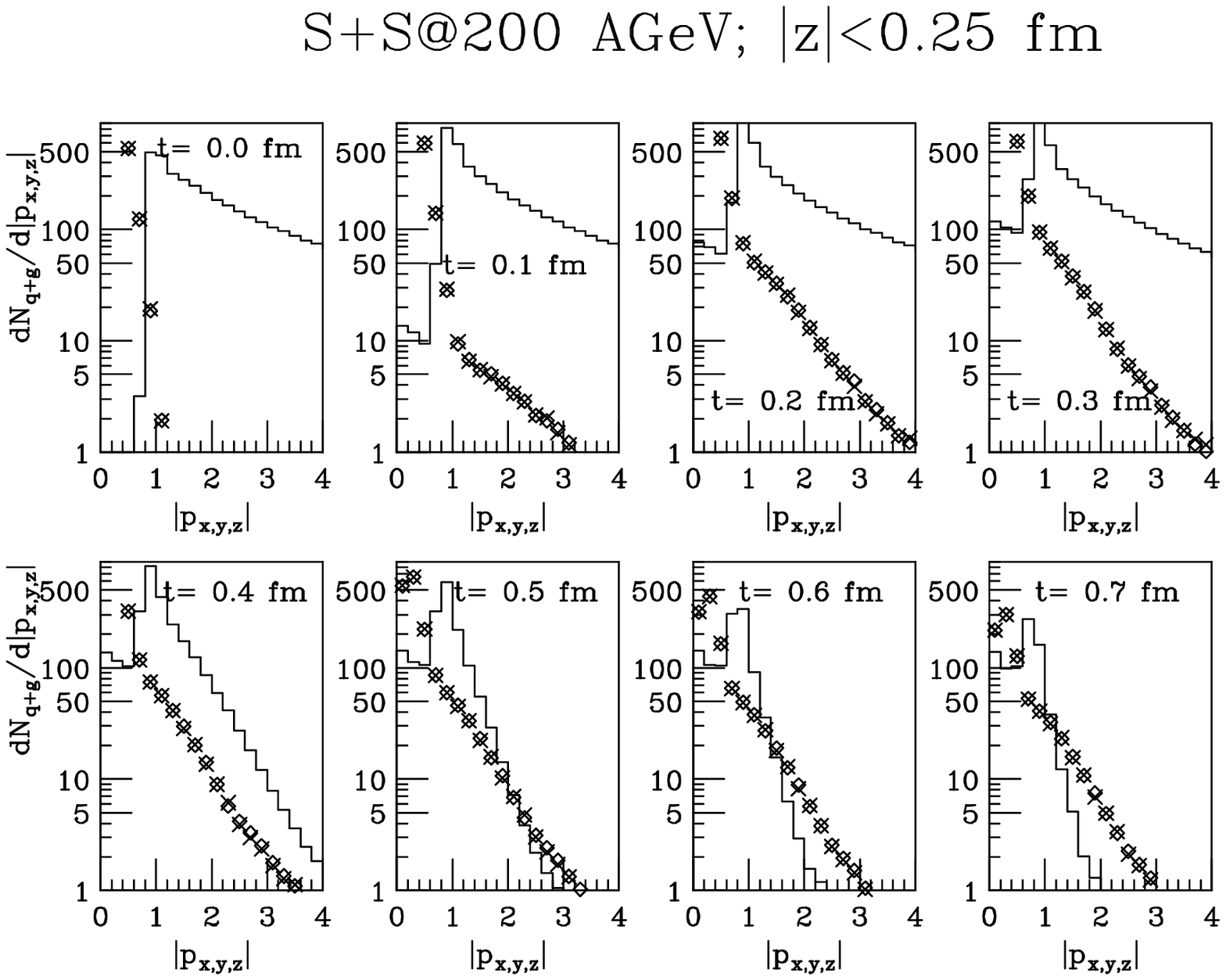,height=12cm,width=15cm}
\vskip 0.1in
\caption{
        Time evolution of the $|p_x|$ (crosses), $|p_y|$ (diamonds)
and $|p_z|$ (histogram) distribution of the (real) partons in 
central collision of sulfur nuclei at $\sqrt{s}=$  200 A$\cdot$GeV.
}
\end{figure}
\newpage
\begin{figure}
\psfig{file=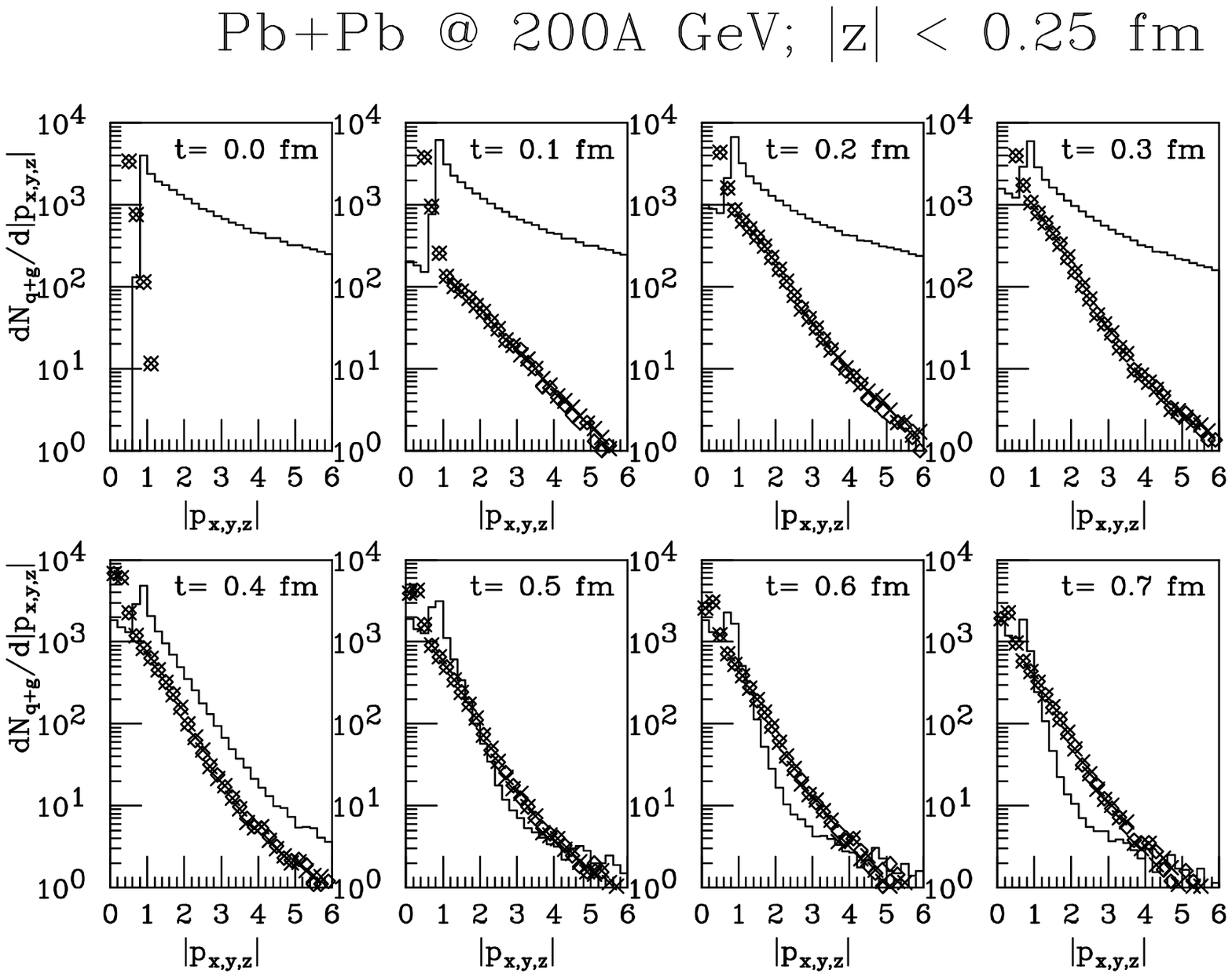,height=12cm,width=15cm}
\vskip 0.1in
\caption{
        Time evolution of the $|p_x|$ (crosses), $|p_y|$ (diamonds)
and $|p_z|$ (histogram) distribution of the (real) partons in 
central collision of lead nuclei at $\sqrt{s}=$  200 A$\cdot$GeV.
}
\end{figure}

\bigskip
 
\begin{thebibliography}{37}

\bibitem{jaipur} Physics and Astrophysics of Quark-Gluon Plasma, Ed.
B. C. Sinha, D. K. Srivastava, and Y. P. Viyogi, Narosa Publishing
House, New Delhi, 1998.

\bibitem{stock} R. Stock, eprint hep-ph/9901415.

\bibitem{gm} K. Geiger and B. M\"uller, Nucl. Phys. B {\bf 369}, 600 (1992).


\bibitem{all}
K. Geiger, Phys. Rep. {\bf 258}, 376 (1995) and references there-in.

\bibitem{vni} K. Geiger, Comp. Phys. Comm. {\bf 104}, 70 (1997);
K. Geiger, R. Longacre , and D. K. Srivastava; nucl-th/9806102, 
The results reported here have been obtained using the
version 4.09, with the removal of the incorrect rescaling of
the space-time co-ordinates. This (incorrect) rescaling was introduced
during the transition of the version 4.08 to 4.09. We are grateful to
S. Bass for drawing our attention to this.

\bibitem{sps} K. Geiger and D. K. Srivastava, Phys. Rev. C {\bf 56},
2718 (1997);
 D. K. Srivastava and K. Geiger, Phys. Lett.  B {\bf 422}, 39 (1998);
 K. Geiger, Nucl. Phys.  A {\bf 638}, 551c (1998);
K. Geiger and B. M\"{u}ller, Heavy Ion Physics {\bf 7}, 207 (1998);
D. K. Srivastava and K. Geiger, nucl-th/9808042.


\bibitem{other}
D. K. Srivastava and K. Geiger, Nucl. Phys. A {\bf 647}, 136 (1999);
D. K. Srivastava and K. Geiger, Phys.  Rev. C {\bf 58},1734 (1998); 
D. K. Srivastava,  nucl-th/9901043.


\bibitem{eg}
             J. Ellis and K. Geiger, Phys. Rev. D {\bf 52}, 1500 (1995);
 J. Ellis and K. Geiger, Phys. Rev. D {\bf 54}, 1967 (1996).

\bibitem{thermal}R. C. Hwa and K. Kajantie, Phys. Rev. Lett. {\bf
56}, 637 (1986); J.-P. Blaizot and A. H. Mueller, Nucl. Phys. B {\bf
289}, 847 (1987); E. Shuryak, Phys. Rev. Lett. {\bf 68}, 3270 (1992);
T. S. Biro, E. van Doorn, B. M\"{u}ller, M. H. Thoma, and X.-N. Wang,
Phys. Rev. C {\bf 48}, 1275 (1993).

\bibitem{kari} K. J. Eskola and X.-N. Wang, Phys. Rev. D {\bf 49}, 1284
(1994).

\bibitem{mik}  M. Gyulassy and P.  Levai, Phys. Lett. B {\bf 442},1 (1998).

\bibitem{bm} B. M\"{u}ller, nucl-th/9902065.

\bibitem{sspc}
K. J. Eskola, B. M\"{u}ller, and X.-N. Wang, Phys. Lett. B {\bf 374},
20 (1996).

\bibitem{smm} 
D. K. Srivastava, M. G. Mustafa, and B. M\"{u}ller, Phys. Lett. B {\bf 396},
 45 (1997); D. K. Srivastava, M. G. Mustafa, and B. M\"{u}ller,
 Phys. Rev. C {\bf 56}, 1064 (1997).

\end{thebibliography}
\end{document}